# Wireless LAN: Past, Present, and Future

Keith Holt
Intel Corporation

## Abstract

*This paper retraces the historical development of wireless LAN technology in the context of the pursuit of ever higher data rate, describes the significant technical breakthroughs that are now occurring, and speculates on future directions that the technology may take over the remainder of the decade. The challenges that these developments have created for low power operation are considered, as well as some of the opportunities that are presented to mitigate them. The importance of MIMO as an emerging technology for 802.11 is specifically highlighted, both in terms of the significant increase in data rate and range that it enables as well as the considerable challenge that it presents for the development of low power wireless LAN products.*

## Introduction

Wireless LAN technology is evolving at a rapid pace. Within just a few years the industry has seen the highest data rates provided by products based on the 802.11 standards migrate from 2 Mbps (802.11) to 11 Mbps (802.11b) and now to 54 Mbps (802.11a/g). Currently, active efforts are underway within the IEEE to define a next generation standard to be known as 802.11n that will enable rates potentially as high as 600 Mbps in a 40 MHz channel. In addition, efforts are also underway to define the 802.11s standard for mesh networking. These developments, as well as potential future technologies such as cooperative diversity, are creating new challenges and opportunities in the design of low power wireless LAN products.

## Historical Developments

The development of the air interface for the earliest 802.11 standard was largely driven by regulatory constraints concerning the use of unlicensed spectrum. Rules established in the U.S. by the Federal Communications Commission (FCC) were written primarily to ensure fair and equal access by mandating a certain level of robustness to interference via spread spectrum techniques. The spectral efficiency of the early 802.11 standard was thus rather limited, realizing only 0.1 bps/Hz with a maximum data rate of 2 Mbps in a 20 MHz channel.

Both direct-sequence (DSSS) and frequency hopping (FHSS) forms of spread spectrum were standardized as alternative means of complying with the mandated 10 dB processing gain requirement.

In the development of the 802.11b standard, the maximum data rate was increased to 11 Mbps as the 10 dB processing gain requirement was relaxed in lieu of the transmitted signal maintaining certain essential properties characteristic of a DSSS system. In 802.11b, a combined modulation and coding scheme known as CCK was adopted to increase rate while maintaining a DSSS like signature to other users of the unlicensed band. With CCK, a spectral efficiency of 0.5 bps/Hz was achieved, representing a fivefold increase over the earlier standard.

The large commercial success of wireless LAN products based on these early standards motivated regulatory bodies in many countries around the world to open additional spectrum at 5 GHz for unlicensed use. This time, however, requirements for bandwidth expansion or spreading as a means for resource sharing were sidestepped in favor of allowing more spectrally efficient modulation schemes for support of high data rate applications.

In the 802.11a standard, OFDM was adopted as the means for achieving a wideband spectrally efficient modulation. A maximum data rate of 54 Mbps yielded a spectral efficiency of 2.7 bps/Hz, representing yet again an approximately fivefold increase over the previous standard. With additional regulatory changes, the same OFDM technology was allowed into the 2.4 GHz band and was standardized as 802.11g.

With regulatory barriers to higher spectral efficiency now removed, higher data rates have become limited more fundamentally by technology, not government policy. In fact, the spectral efficiencies achieved by the OFDM based standards were quite good, and essentially represented the best that could be achieved within the practical constraints of cost and range.

## Emerging Developments

Fundamental breakthroughs in information theory, which first emerged during the time of the early development of wireless LANs, have now reached a level of maturity and acceptance that is allowing them to drive the quest for higher spectral efficiencies and data rates. So called "MIMO", or multiple-input, multiple-output antenna technology allows spectral efficiencies and hence data rates which were heretofore unreachable. The future 802.11n standard is certain to incorporate this technology, and efficiencies up to 15 bps/Hz are likely to be specified at the highest rate modes which maintains the historical trend of fivefold increases with each new standard.

MIMO technology will provide benefits beyond only enhanced data rate. Through the availability of spatial diversity provided by multiple antennas, the range of a wireless LAN network in a fading multipath environment is extended several-fold relative to a conventional signal antenna or SISO system. Other likely enhancements in the 802.11n standard will also increase the range of wireless networks, such as the use of LDPC codes. Even closed loop, transmit side beamforming may be specified in order to improve rate and reach.

Another important development in wireless LAN technology is the emergence of mesh networking. Mesh networks have the potential to dramatically increase the area served by a wireless network. Mesh networks even have the potential, with sufficiently intelligent routing algorithms, to boost overall spectral efficiencies attained by selecting multiple hops over high capacity links rather than single hops over low capacity links.

## Future Developments

One possible avenue of future development in wireless LAN technology is in the area of "cooperative diversity." Cooperative diversity can be viewed as somewhat of a cross between MIMO techniques and mesh networking. In a cooperative diversity scheme, redundancy in transmission is achieved in a manner analogous with diversity transmission in MIMO. However, the redundant transmission is realized via the cooperation of third party devices rather than solely from the originating device. In a cooperative diversity scheme, third parties which can successfully decode an on-going exchange will effectively regenerate and relay, with appropriate coding, the original transmission in order to improve the effective link quality between the intended parties.

## Low Power

These past, present, and future developments create many challenges in the development of low power wireless LAN devices. For example, beginning with the introduction of OFDM, the high peak-to-average ratios characteristic of spectrally efficient modulation have resulted in low power efficiency of the power amplifier and other components in order to achieve the necessary high linearity.

Emerging MIMO technologies create whole new challenges for designers striving to achieve power efficient operation. Multiple transmit and receive RF chains, not to mention the additional baseband processing involved, significantly increase the power consumption over single antenna devices. Advanced coding schemes and increased baseband processing required for beamforming solutions will place additional power demands on future devices.

At the same time, some of these new technologies also provide opportunities to mitigate their increased power appetite through clever product implementation. For example, MIMO systems could reduce power by switching off all but one receive chain until a packet is detected, switching on the additional chains only as required to decode high rate traffic. Closed loop beamforming techniques could allow for effective transmit power control. Finally, mesh or cooperative diversity schemes could "share" some of the power burden with willing third party devices that are less power constrained, such as a device that is drawing power from an electrical outlet rather than a battery.

It is also important that future wireless LAN protocol standards be designed from the outset for low power operation. In computer notebooks, wireless power consumption represents only a fraction of the overall platform power budget. On the other hand, smaller form factor devices impose more stringent power requirements. Wireless LAN protocols currently make few concessions to issues of power management as compared to cellular air interface standards. Undoubtedly, future wireless LAN standards could benefit from more attention in this area.

## Conclusions

Wireless LAN technology has experienced extraordinary advances in rate, range, and spectral efficiency. These advances, initially constrained by regulatory policy related to the use of unlicensed spectrum, are now driven entirely by technical innovation and end user application demand which show no sign of abatement. As these new technical innovations continue, the challenges of providing low power operation will continue to mount while new opportunities to lower power consumption will present themselves to innovative designers.